\documentstyle[12pt,amstex,amssymb] {article}
\newcommand{\beq}{\begin{equation}}

\newcommand{\eeq}{\end{equation}}
\newcommand{\beqa}{\begin{eqnarray}}
\newcommand{\eeqa}{\end{eqnarray}}
\begin {document}
\parindent=15pt
\begin{flushright}
{\bf US-FT/16-97}
\end{flushright}
\vskip .8 truecm
\begin{center}
{\bf PRODUCTION ASSOCIATED TO RARE EVENTS IN HIGH ENERGY HADRON-HADRON
COLLISIONS.}\\
\vskip 1.5 truecm
{\bf J. Dias de Deus$^*$, C. Pajares and C. A. Salgado.}\\
\vskip 0.9 truecm
{\it Departamento de Fisica de Part\'{\i}culas, Universidade de Santiago de
Compostela, \\
15706--Santiago de Compostela, Spain}
\end{center}
\vskip 2.5 truecm

\begin{abstract}
\hspace {30pt}

At very high energy the same universal relation between the multiparticle or
the transverse energy distribution associated to a rare event $C$, $P_C$ and
the corresponding minimum bias distribution P, $P_C(\nu)\equiv \nu/<\nu>
P(\nu)$, $\nu\equiv n$ or $E_T$ works for nucleus-nucleus collisions as well as
for hadron-hadron collisions. This suggests that asymptotically, all hadronic
processes are similar.
\end{abstract}

*) Iberdrola visiting professor. On leave of absence from 
Instituto Superior T\'ecnico, 1096 LISBOA codex, Portugal.

\vskip 1cm
May 1997 \\
\vskip .2 truecm

{\bf US-FT/16-97}

\pagebreak

Successful models for multiparticle production in hadron-hadron, hadron-nucleus
and nucleus-nucleus collisions are multiple interaction models where the final
multiplicity distribution and correlations are obtained from the superposition
of the elementary collisions. A typical model of this kind is the Dual Parton
Model (DPM) \cite{1,2}.

We have recently made an attempt to abstract the general properties of such
models in the limit when the number of elementary collisions is very large
\cite{3,4}. There are two possibilities for increasing the number of elementary
collisions. One is by increasing the number of constituents, i.e., by
increasing the atomic number of the incoming particles and going to heavy ion
collisions. The other possibility is by increasing the center of mass energy and
going to hadron colliders. It is this possibility that we would like to
discuss here.

In models where particle distributions are obtained from the superposition of
independent contributions of the elementary collisions one can in general,
write

\begin{equation}
P(n)=\sum_{\nu=1}\sum_{n_1,\dots,n_\nu} \varphi(\nu)p(n_1)p(n_2)\dots
p(n_\nu)
\label{1}
\end{equation}

\noindent
where $P(n)$ is the normalized multiplicity distribution , $\varphi(\nu)$ the
distribution in the number of elementary collisions and $p(n_i)$ the elementary
collision particle distribution. A strong approximation is made in (\ref{1}):
all elementary collisions are treated as equivalent.

From (\ref{1}) it follows, in a straightforward manner, for the average
multiplicity $<n>$ and the square of the width of the 
distribution $D^2\equiv <n^2>-<n>^2$
\cite{3}
\begin{equation}
<n>=<\nu>\bar n
\label{2}
\end{equation}

\noindent
and
\begin{equation}
{{D^2}\over{<n>^2}}={{<\nu^2>-<\nu>^2}\over{<\nu>^2}}+{{1}\over{<\nu>}}
{{d^2}\over{\bar n^2}}\ \ ,
\label{3}
\end{equation}

\noindent
where $\bar n$ and $d^2$ are the quantities equivalent to $<n>$ and $D^2$,
respectively, for the elementary collision. Relations similar to (\ref{3}) can
be written for higher order correlations \cite{3}.

It is experimentally known that the average multiplicity $<n>$ and the
normalized dispersion $D/<n>$ increase with the atomic number A of the
colliding particles and with the energy. In particular, $D/<n>$ is larger in AA
collisions than in hh collisions and increases with energy as shown by KNO
scaling \cite{5} violations at the CERN $p\bar p$ collider, \cite{6}.

These observations imply, from (\ref{2}) and (\ref{3}), that $<\nu>$ and
$(<\nu^2>-<\nu>^2)/$ $<\nu>^2$ increase with energy (and A). Asymptotically we
obtain 
\begin{equation}
{D^2\over<n>^2}={<\nu^2>-<\nu>^2\over <\nu>^2 }
\label{4}
\end{equation}

In general, at high energy, fluctuations in the number of elementary
collisions will dominate correlations.

A word of caution is needed here. At present $p\bar p$ collider energies,
(0.2-2) TeV, the average number $<\nu>$ of elementary collisions is not
large \cite{2}.
On the other hand, valence quark-valence quark interactions are still
dominant making the approximation "all elementary collisions are treated as
equivalent" somehow suspicious. This is particularly true if one triggers on
fast particles ($x\simeq 1$) or low $p_T$ heavy weak bosons ($W$ or $Z^0$):
they must come from valence quarks.

We look now at the multiparticle distribution $P_C$ associated to a rare weakly
absorbed event C, in comparison with the minimum bias distribution P. The
discussion of \cite{4} on nucleus-nucleus collisions, extended to high energy
hadron-hadron collisions, means that the universal relation 
\begin{equation}
P_C(n)={n\over<n>}P(n)
\label{5}
\end{equation}

\noindent
should hold. This relation can as well be written for transverse energy $E_T$
distributions.

Eq. (\ref{5}) was shown to work reasonably well in nucleus-nucleus collisions.
Before testing (\ref{5}) for high energy hadron collisions we would like first
to further discuss its content.

From (\ref{5}) one arrives at the following relations for moments $<n^q>$:

\begin{equation}
<n^q>_c=<n^{q+1}>/<n>, \ \ \ \ q=1,2,\dots
\label{6}
\end{equation}

\noindent
with, in particular,
\begin{equation}
<n>_c={<n^2>\over<n>^2}<n>,
\label{7}
\end{equation}

\noindent
or,
\begin{equation}
<n>_c\ge <n>.
\label{8}
\end{equation}

\noindent
Making use of the KNO variable,
\begin{equation}
z\equiv n/<n>
\label{9}
\end{equation}

\noindent
(\ref{6}) becomes
\begin{equation}
<z^q>_c=<z^{q+1}>/<z^2>^q.
\label{10}
\end{equation}

If one now writes the probability distribution in the KNO form,
\begin{equation}
<n>P(n)\longrightarrow \Psi(z),
\label{11}
\end{equation}

\noindent
with $\int \Psi(z)dz=\int z\Psi(z)dz=1$, one sees that the event C KNO function
goes to zero as $z\to 0$, faster than the standard KNO function, because of the
factor $n$ in (\ref{5}), and that, at large $z$, as $<n>_c><n>$, it is
displaced to smaller values of $z$.

Parametrizing the function $\Psi(z)$ by using the generalized gamma function,
with parameters $\kappa$ and $\mu$, $\kappa>0$ and $\mu>0$,

\begin{equation}
\Psi(z)={\mu\over\Gamma(\kappa)}\Big[{\Gamma(\kappa+1/\mu)\over\Gamma
(\kappa)}\Big]^{\kappa\mu}z^{\kappa\mu-1}exp\Big(-\Big({\Gamma(\kappa
+1/\mu)\over\Gamma(\kappa)}z\Big)^\mu\Big),
\label{12}
\end{equation}

\noindent
with

\begin{equation}
<z^q>={\Gamma(\kappa+q/\mu)\over\Gamma(\kappa+1/\mu)^q}\Gamma(\kappa)^{q-1},
\label{13}
\end{equation}

\noindent
from (\ref{10}) one easily sees that if the parameters of the minimum bias
multiplicity distribution are $\kappa$ and $\mu$, the parameters $\kappa_C$ and
$\mu_C$ of the distribution associated to event C are related to $\kappa$ and
$\mu$ by:

\begin{equation}
\kappa_c=\kappa+{1\over\mu},
\label{14}
\end{equation}
$$
\mu_c=\mu.
$$

Let us now consider the application of (\ref{5}) and (\ref{10}) to $p\bar p$
collisions at CERN and Fermilab collider energies.


\vskip 0.5cm
{\bf 1 Multiplicity distributions associated to $W^\pm$, $Z^0$ production
($~\protect\sqrt{s}=1.8$ TeV)}
\vskip 0.5cm

The CDF Collaboration at Fermilab has measured the multiplicity (and $E_T$)
distribution associated to $W^{\pm}$ and $Z^0$ production in high energy $p\bar
p$ collisions \cite{8}. For large enough $p_T^W$($\gtrsim 5$ GeV/c) $<n>_c$ 
becomes independent of $p_T^W$ as expected. For smaller values of $p_T^W$,
$<n>_c$ approaches $<n>$ and the distributions become identical. This can be
understood in the following way: as triggering in low $p_T$ $W$ or $Z^0$ 
means triggering on valence quark collisions, and these collisions are always
present, such triggering is not selective (gives the same result as without
trigger). The relevance of valence quarks is reflected also in the observed
forward-backward $W^+-W^-$ asymmetry. At high $p_T$ hard collisions, Drell-Yan
$W$ and $Z^0$ production, low $x$ parton collisions become important and the
rare event probability becomes proportional to $\nu$.

In Fig.1 we directly test (\ref{5}) by using CDF data. The result is fairly
reasonable. Note that the existing $E_T$ distributions \cite{8} are also
consistent with (\ref{5})

Associated particle  production to large $E_T$ jets and $W^\pm$, $Z^0$
detection has been previously discussed, in the framework of DPM, in \cite{9}.

\vskip 0.5cm
{\bf 2 Multiparticle distributions associated to jet events
($~\protect\sqrt{s}=0.2-0.9$ TeV)}
\vskip 0.5cm

The UA1 Collaboration at CERN has studied multiplicity and $E_T$ distributions
associated to jet events, characterized by $E_T^{jet}> 1.5$ GeV in the
acceptance region $|\eta|<1.5$ and $\Delta\phi< 30^0$ from the vertical, and
compared the obtained KNO distribution to the no-jet event distribution
\cite{10,11}. It was found that the average multiplicity and the transverse
energy density of events with jets was larger (by a factor of the order of 2)
than in the no-jet events and both quantities were independent of $E_T^{jet}$,
$E_T^{jet}>5$ GeV.

This kind of behavior is expected in our approach. For rare events the
associated $<n>$ and $<E_T>$ are larger, see (\ref{10}), than in the remaining
events. As far as the jet criterion does not allow for too many events
($E_T^{jet}>5$ GeV seems to be the right criterion) the associated average
multiplicity and transverse energy should be independent of $E_T$.

This last point is true at not very high energies, $\sqrt{s}$=0.2 GeV, where the
measured jet cross section is 2.4 mb \cite{10}. This cross section rises very
fast as the energy increases. Therefore at higher energies the jet events are
not rare and they are strongly absorbed. For this reason we compare our
prediction only with the experimental data at $\sqrt{s}$=0.2 TeV.

However, we notice that from the only data available on the KNO no-jet and jet
associated multiplicity function measured in the pseudorapidity range
$|\eta|<2.5$ and from the quoted no-jet and jet associated multiplicity
$<n>_{jet}=26.5\pm 0.2$, $<n>_{no-jet}=13.8\pm0.1$, the obtained minimum bias
multiplicity distribution through the formula:

\begin{equation}
P(n)=P_{jet}(n){\sigma^{jet}\over\sigma^{in}}+P_{no-jet}(n)
{\sigma^{no-jet}\over\sigma^{in}}
\label{nova}
\end{equation}

\noindent
and $\sigma^{jet}=2.4 mb$, $\sigma^{in}=40 mb$, $\sigma^{no-jet}=37.6 mb$ does
not coincide with the experimental minimum bias data measured in the same
pseudorapidity range as it is seen in Fig.2. It is seen that for $n>35$ there
is a clear disagreement which could be arranged by means of a broader jet
distribution. Notice that a larger fraction $\sigma^{jet}/\sigma^{in}$ would
improve the agreement at large $n$ but now the disagreement will appear for $n$
arround 30.

Due to that, it is not expected a good agreement between our prediction and the
actual experimental data and this is seen in Fig.3, where we show our prediction
together with the jet associated multiplicity distribution obtained from the
experimental data on the corresponding KNO function using the jet associated
multiplicity obtained from (\ref{7}). From Fig.3 it is seen that in order to
obtain a good agreement, the experimental jet associated multiplicity
distribution should be a little broader. We remark that this conclusion is the
same that we deduced above by direct comparison with experimental data.

\vskip 0.5cm
{\bf 3 Multiplicity associated to annihilation events}
\vskip 0.5cm

The annihilation events in $p\bar p$ 
collisions are also rare events which should
satisfy \cite{5}. In fact, they are shadowed by themselves and as far as their
cross section is small, the consequent absorption is also small \cite{13,14}.
Unfortunately, there is not data on topological annihilation cross section at
high energy \cite{15} and for this reason the comparison is at $p_{lab}=7$
GeV/c. In Fig.4 it is show the minimum bias $\bar pp$ multiplicity distribution
together the annihilation data \cite{15} and our result which is again fairly
reasonable.

\vskip 0.5cm

As a conclusion we may say that even in $\bar pp$ collision, the simple
approach developed and tested for nucleus-nucleus collisions in \cite{3} and
\cite{4} can also be applied, which shows the same physics. 

It can be shown
that in the limit of infinite energy and therefore infinite multiplicity the
multiplicity distribution of $\bar pp$ and nucleus-nucleus collisions is
controlled by the multiplicity distribution of a single collision \cite{16}.
However, at the present energies the multiplicity distribution 
is controlled by the distributions on the number of collisions.

This fact can only be compatible with the infinite energy result if there is
some transition mechanism which transforms 
the multiple scattering structure into a
single one. A mechanism of this type is the percolation of strings \cite{17}.
As the density of strings increases due to the increasing of energy or to the
size of the collision objects, they begin to overlap each other . When a
critical density is reached, the percolation threshold, it is possible to go
through the whole collision surface by a path of overlapping strings. In this
way, the effective number of collisions is reduced drastically. The percolation
of strings has been proposed to explain the recent $J/\Psi$ suppression
\cite{18} found by the NA50 Collaboration in Pb-Pb collisions \cite{19,20} 
and can
be seen as the way of taking place the phase transition from hadronic matter to
quark gluon plasma.
 
Acknowledgements

This work has been done under the contract AEN96-1673 of CICYT of Spain. We
thank Iberdrola for financial support and one of us (CAS) to Xunta de Galicia
for a fellowship. We thank M. Braun for discussions, in particular on the
asymptotic behavior of the multiplicity distributions.

\pagebreak

\newpage

\begin{center}
{\bf Figure captions}\\
\end{center}
\vskip 1cm
\noindent
{\bf Fig.1}. Prediction for the associated multiplicity distribution for
$W^{\pm}$ and $Z^0$ events (round black points) together with the experimental
data ($W^{\pm}$ cross points and $Z^0$ white stars) and the minimum bias
distribution (squares). 

\vskip 1cm
\noindent
{\bf Fig.2}. Minimum bias distribution obtained from equation (\ref{nova}) of
the text (write crosses) and the experimental minimum bias (white squared
points) together with the jet (black stars points) and no-jet (white rounds)
associated multiplicity distribution. 

\vskip 1cm
\noindent
{\bf Fig.3}. Prediction for the jet associated multiplicity distribution (black
round points) together with the experimental data and minimum bias multiplicity
distribution (white squared points). 

\vskip 1cm
\noindent
{\bf Fig.4}. Prediction for the KNO function for annihilation events (dashed
curve) and the experimental data (stars) together with the minimum bias data
(white squared points) parametrized as (\ref{12}) (solid line).

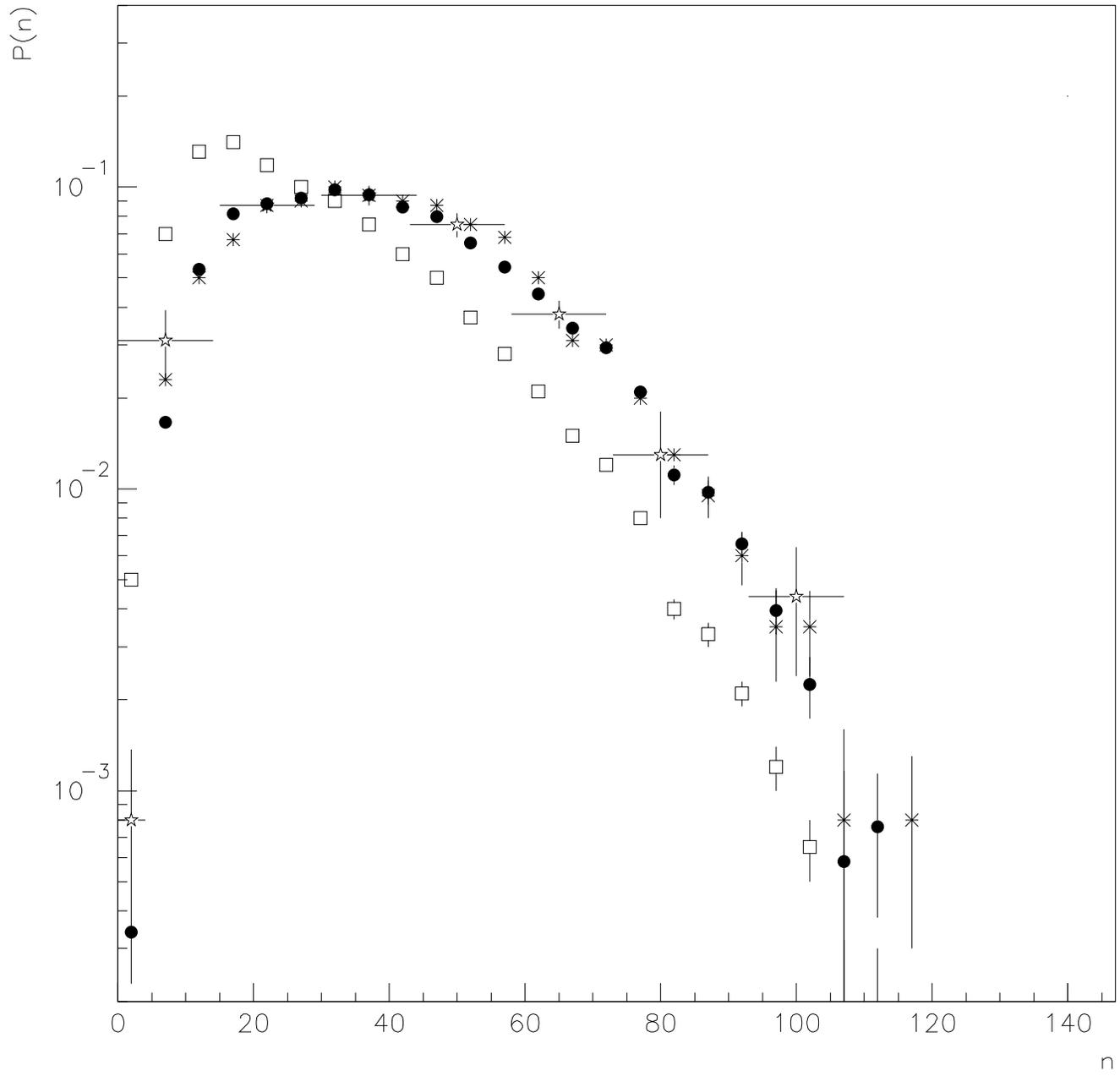
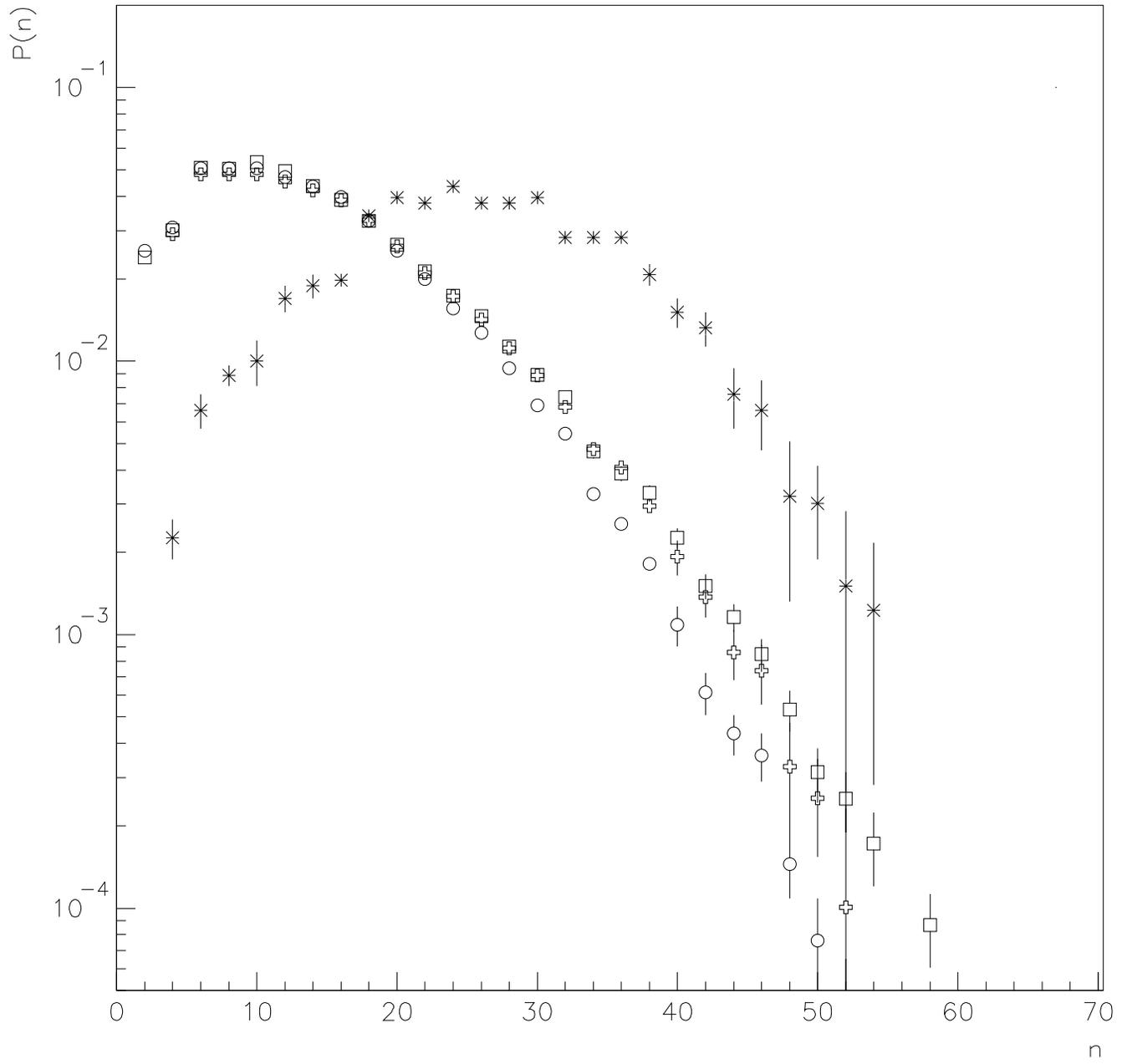
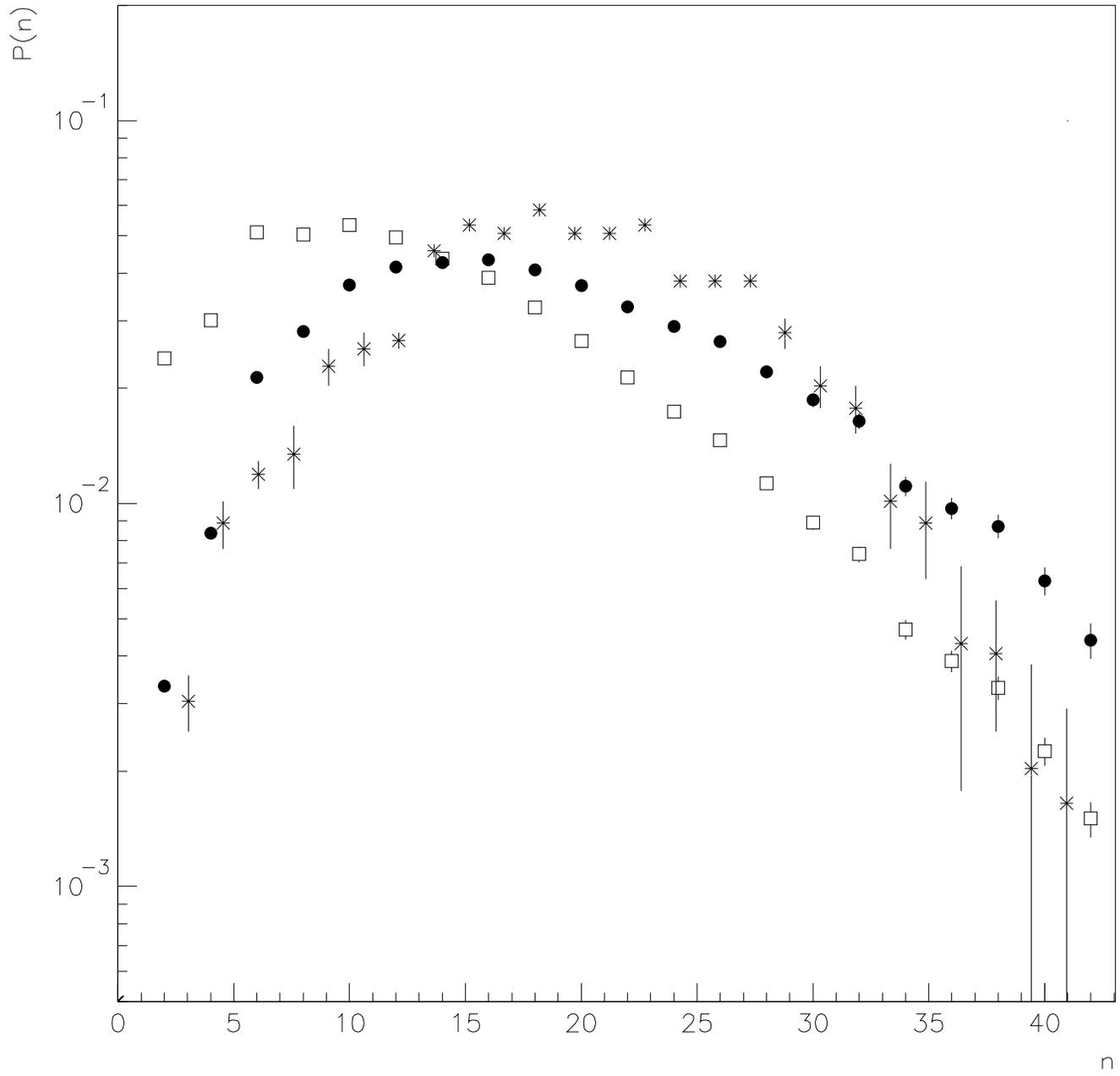
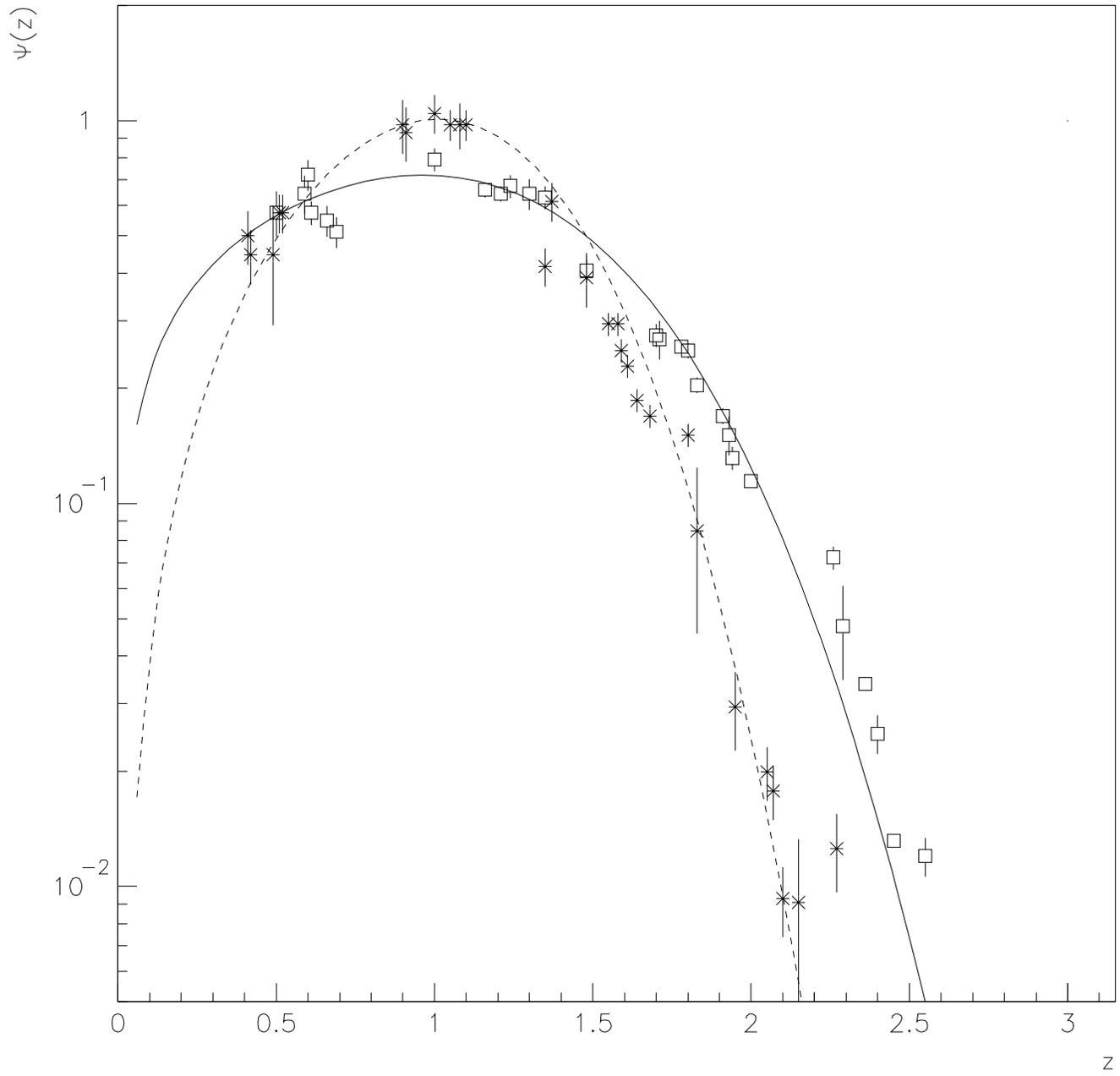
\end{document}